\documentclass[10pt]{article}
\usepackage[utf8]{inputenc}
\usepackage{amsmath,amsfonts,amssymb,latexsym}
\usepackage[T1]{fontenc}
\newtheorem{satz}{Theorem}[section]

\newtheorem{assumption}[satz]{Assumption}
\newtheorem{defi}[satz]{Definition}

\newtheorem{bem}[satz]{Remark}
\newtheorem{lemma}[satz]{Lemma}
\newtheorem{koro}[satz]{Corollary}
\newtheorem{conclusion}[satz]{Conclusion}
\newtheorem{ob}[satz]{Observation}

\newtheorem{conjecture}[satz]{Conjecture}

\newcommand{\mcal}{\mathcal}
\newcommand{\mbf}{\mathbf}

\newcommand{\tit}{\textit}

\newcommand{\beq}{\begin{equation}}
\newcommand{\eeq}{\end{equation}}

\begin{document}
\thispagestyle{empty}
\begin{center}
\vspace*{1.0cm}
{\Large{\bf The Black Hole Singularity as a Thermodynamic System being the Seat of BH Entropy}}
\vskip 1.5cm
 
{\large{\bf Manfred Requardt}}

\vskip 0.5cm

Institut fuer Theoretische Physik\\
Universitaet Goettingen\\
Friedrich-Hund-Platz 1\\
37077 Goettingen \quad Germany\\
(E-mail: muw.requardt@googlemail.com) 

\end{center}
\begin{abstract}
We argue in this paper that the entropy of the BH is located in the BH singularity and that the localization around the event horizon is perhaps a secondary effect. We show in particular that the dependence of the entropy located in the singularity is also proportional to $M^2$. We furthermore show that our analysis leads in a natural way to information loss as the micro structure of the BH singularity is of a completely different character compared to ordinary quantum space time underlying our classical space time manifold.
\end{abstract}
\newpage
\section{Introduction}
Papers about the socalled black hole (BH) information paradox go presumably into the hundreds if not thousands since Bekenstein and Hawking wrote their meanwhile classic papers (\cite{Bek1},\cite{Hawk1},\cite{Bek2},\cite{Hawk2}. Bekenstein associated the surface area of the event horizon of a BH with the thermodynamic entropy of the BH, thus establishing it as a thermal object while, among other things, Hawking shows with the help of a semiclassical analysis that a BH emits a thermal radiation.
\begin{bem}For convenience we discuss in this paper only a BH with parameter its mass $M$.
\end{bem}

Bekenstein formulated a \tit{generalized second law}
\beq \Delta S_{tot}:= \Delta S_{BH}+ \Delta S_{ord} \geq 0   \eeq
with $S_{ord}$ denoting some ordinary entropy located exterior to the BH and $S_{BH}$ being the entropy of the BH which reads (in SI units, with $k_B=1$)
\beq S_{BH}=Ac^3/4\hbar G=A/4L^2_p    \eeq
$A$ denoting the surface area of the event horizon
\beq A_{BH}=16\pi(GM/c^2)^2  \eeq
and $L_p$ the Planck length
\beq  L_p=\sqrt{G\hbar/c^3}  \eeq
In Planck units ($G=c=\hbar=1$) we have simply
\beq S'_{BH}=A'/4\quad\text{with}\quad A=A'\cdot L_p^2  \eeq

The socalled \tit{Hawking temperature}, measured at spatial infinity, is (in SI units)
\beq T_{\infty}=\hbar c^3/8\pi Gm=T'\cdot T_p=(8\pi M')^{-1}\cdot \hbar c^3/G\sqrt{\hbar c/G}  \eeq
With $M_p=\sqrt{\hbar c/G}$ the Planck mass and $T_p=\sqrt{\hbar c^5/G}$ the Planck temperature, this yields
\beq T'=(8\pi M')^{-1}   \eeq

The socalled \tit{BH information paradox} consists of the following problem: Hawkings \tit{semiclassical} calculation shows that the emitted radiation is exactly thermal. This does not present a problem as long as there do exist enough states inside the BH horizon since \tit{entanglement} between internal and external quantum states should be quite natural. However, after the BH has completely evaporated what remains is a thermal radiation state with nothing left to couple to. Consequently, a pure initial state has evolved into a mixed thermal state.
\begin{bem}In the next section we will discuss in more detail whether it is reasonable to work with pure states in this context.
\end{bem}
Such a scenario is impossible in ordinary quantum theory! This problem has been discussed extensively with quite a few resolutions being proposed by some authors which then were frequently criticized by other authors but a univerally accepted solution does not have emerged in our view.

As we feel unable to give a complete review of this quite varied situation we will only cite a few (as we think) representative papers. Most of the authors argue that \tit{information} cannot be lost in physics. This vague statement is frequently regarded as a fundamental physical law, while we are unable to find a valid argument in favor of this statement. What does only hold in our view is \tit{unitarity} in ordinary standard quantum theory. If one however is willing to accept this assumption, then one is forced to argue in favor of the existence of quite contrived and long range correlations or entanglement between degrees of freedom (DoF) inside and outside of the BH event horizon and at different times. This then may lead to problems like the \tit{firewall paradox} (see \cite{fire}). Some older papers are for example \cite{Harlow},\cite{Susskind},\cite{Preskill},\cite{Page},\cite{Carlip},\cite{Huijse},\cite{Lowe1},\cite{Lowe2},\cite{Lowe3},\cite{Giddings1},\cite{Giddings2},\cite{Giddings3},\cite{Mathur},\cite{Jacobson}.

More recent papers, which employ a variety of fairly advanced methods, but which also rely on the lessons coming from the \tit{ADS-CFT conjecture}, are for example \cite{Alm},\cite{Raju} or \cite{BoussoSnow}. One should however note that the \tit{ADS-CFT conjecture} depends crucially on the \tit{bulk-boundary scenario} which is typical for \tit{hyperbolic geometries} and does not seem to prevail in the space-time context we are generally living in. This is for example exemplified in the beautiful book \cite{Ratcliffe}, where the rigid connection between symmetries on the boundary and the bulk of hyperbolic geometries is discussed in quite some detail.

Yet another approach is developed in e.g. \cite{Requ1} where we construct a model in which quantum matter (QM) degrees of freedom (DoF) and quantum gravitational DoF happen to be explicitly entangled and which remains so even after the complete evaporation of the BH.

What all these approaches have in common is their focus on the region of the BH event horizon and practically everything what is needed to derive the unitarity of the evolution is assumed to be located there or in its neighborhood. We surmise that the main reason for that is the observation that the entropy happens to be proportional to the BH horizon which is viewed as the primary phenomenon.

While it is perhaps fair to say that the majority of experts in the field believe in a unitary evaporation process, thus believing in the universality of orthodox quantum theory, there are however a few notable exceptions apart from Hawking himself. In \cite{Wald1},\cite{Wald2},\cite{Unruh1} it is speculated that  the BH singularity may perhaps play a greater role which is possibly overlooked in the majority of approaches coming mainly from particle theorists with their strong belief in the universality of unitary quantum theory. Furthermore they argue that unitarity may be broken in the evaporation process.

In the following we will argue that the observation that the BH entropy seems to be proportional to the area of the event horizon is, in our view, only a secondary effect and that the entropy is rather being located in the BH singularity, or, as another possibility, it happens to be observer dependent whether the entropy is located in the singularity or around the horizon, or perhaps, in both regions.
\begin{bem}We show in particular that the entropy, assumed to be located in the BH singularity, is proportional to $M^2$, $M$ the mass of the BH, that is, it has the same dependence as the entropy, assumed to be located around the BH event horizon.
\end{bem}
 Our analysis will furthermore show that information is completely destroyed in the Bh singularity, which, in our view, is quite natural as the region of the singularity is a zone where ordinary quantum theory presumably does no longer prevail. 
 \begin{bem}The main results are derived in section 4. In the next section we discuss the role of pure or mixed states in this context. In section 3 we develop a model of the micro structure of the BH singularity.
 \end{bem}
\section{The Information Paradox and the Question of Pure or Mixed States}
Without doubt the BH is a non-closed , i.e. open system of many DoF. That is, if one wants to treat it as a quantum system, the natural context should be the regime of (thermal) \tit{many body systems}. In a certain approximation one may view it as a \tit{microcanonical quantum ensemble} in \tit{quantum statistical mechanics}. This field of physics was exactly created to deal with such problems.

On the other hand, the socalled BH information paradox results from treating the BH initial state as a pure quantum state. this is a little bit puzzling, but there are given certain arguments why this appears reasonable. For example, in the nice lecture notes of `t Hooft (\cite{Hooft}) 
\begin{quote}From a physical point of view the distinction between pure states and mixed states for macroscopic objects is pointless. Black Holes should be regarded as being macroscopic. So it is very likely that what we perceive as a mixed state is actually a pure state whose details we were unable to resolve. \end{quote}

A slightly more detailed remark can be found in \cite{BekPure}: \begin{quote}
One can imagine a Black Hole formed from matter in a pure state, e.g. a gravitating sphere of superfluid at $T=0\, K$. The unitarity principle would thus require that the system always remain in a pure state. \end{quote}

While this latter remark adresses the real problems, it is, on the other hand, obvious that practically no BH-process will fulfill this requirements. We will therefore rather discuss the former remarks by `t Hooft as such a point of view one can find frequently in the literature concerning the BH information paradox.

In classical statistical mechanics it is correct that an ensemble like the microcanonical can be thought of as a group of individual systems each occupying a distinct point in the socalled \tit{phase cell} in configuration space. In quantum statistical mechanics this is however in our view grossly incorrect. 
\begin{bem} It is fundamental in quantum theory that (pure) states have to be prepared. Therefore one has to make a distinction between socalled proper and improper mixed states, notions which have been introduced by D'Espagnat (\cite{Espagnat})
\end{bem}

The former concept was introduced by v.Neumann in \cite{Neumann1} or \cite{Neumann0}. That is, if we have instead of a pure state,$\psi$, prepared an ensemble of pure states,$\{\psi_j\}$, with probabilites $\{p_j\}$, the \tit{density matrix} reads
\beq \rho=\sum_j\, p_j\cdot |\psi_j><\psi_j|\quad ,\quad \sum_j\, p_j=1   \eeq
The same situation hold in a measurement process of an observable $A$ with eigenvalues $a_j$ and eigenstates $\psi_j$, that is
\beq \psi=\sum_j\, c_j\cdot\psi_j\quad , \quad p_j=|c_j|^2  \eeq
This may be called the \tit{ignorance situation}, i.e. the individual pure states do exist but are unknown.

Perhaps more important and more typical are scenarios where one cannot assume that the individual states
 $\psi_j$ do exist. This is, for example, the case if the states of e.g. subsystem $I$ are entangled with the states of a subsystem $II$ and we make observations, restricted to subsystem $I$. A typical example is a non-closed system which is entangled with its environment. With e.g. $\Psi$ a pure state of the total system $I\otimes II$ and $A$ an observable restricted to $I$ we can write 
 \beq (\Psi|(A\otimes 1)\Psi)= Tr_{II}(\rho_I\cdot A)   \eeq
where $Tr_{II}$ denotes the \tit{partial trace} with respect to system $I$ and $\rho_I$ is a density matrix with respect to $I$.

But this scenario may as well prevail in quantum statistical mechanics. On the one hand we may have an open system which is entangled to its environment. On the other hand, even if the system happens to be closed and isolated from the environment to some extent, its state will, in our view, never be a proper one as the individual states making up the density matrix are usually never individually prepared pure states.

It is useful to treat the microcanonical ensemble. We assume that its total energy lies in the interval $[E,E+\Delta E]$ with $\Delta E$ sufficiently small. We assume furthermore an orthonormal basis of eigenvectors of the Hamiltonian $\{\psi_j\}$ be given with energies $E_j$ lying in this interval. Hence every allowed (pure) state of the system can be written as
\beq  \psi=\sum_{j=1}^N\, c_j\cdot \psi_j\quad ,\quad \sum_{j=1}^N\,|c_j|^2=1   \eeq
The microcanonical density matrix has the form
\beq  \rho=N^{-1}\cdot\sum_{j=1}^N\, |\psi_j><\psi_j|=N^{-1}\cdot\mbf{1}_E   \eeq
with $\mbf{1}_E$ the unit operator on the subspace $\mcal{H}_E$ spanned by the $\{\psi_j\}$ or, put differently, the projector onto this subspace.

\begin{bem}
It is important to note that this microcanonical density matrix, $\rho$, does not imply that the system is in one of the energy eigenstates $\psi_j$. We can obviously choose another suitable orthonormal basis $\psi'_j$ with
\beq N^{-1}\cdot\sum\,|\psi_j><\psi_j|=N^{-1}\cdot\sum\,|\psi_j'><\psi_j'|=N^{-1}\cdot\mbf{1}_E\eeq
\end{bem} 

We can in fact prove the following useful lemma
\begin{lemma} It holds
\beq Tr\, (\rho\cdot A)= Exp\, (\psi|A\psi)  \eeq
with $Exp$ the probabilistic expectation over a random vector with uniform distribution over the unit sphere of $\mcal{H}_E$, $\{\psi |\,||\psi ||=1\}$,that is, with respect to Haar measure on $S_{2N-1}$.
\end{lemma}
Sketch of proof: Evaluate various integrals like
\beq \int\, \bar{c}_l\cdot c_m\, d\mu=\delta_{lm}\cdot \int\, |c_l|^2\, d\mu=(2N-1)^{-1}  \eeq
using symmetry arguments by expressing the Haar measure on $S_{2N-1}$ in \tit{hyperspherical polar coordinates} (see e.g. \cite{Ratcliffe} or \cite{Jancel}.
\begin{bem} Note that the coefficients $c_l$ play the role of the Euclidean coordinates $x_l$ which are expressed in hyperspherical polar coordinates.
\end{bem}

In this context we want to mention \cite{Lebo1}, referring to the deep paper of v.Neumann (\cite{Neumann2}), or \cite{Lebo2}.
 \begin{bem} In v.Neumanns paper one can find one of the early and highly nontrivial applications of the concentration of measure phenomenon.
 \end{bem}
 \begin{conclusion}The above remarks show clearly that in general the microcanonical ensemble cannot be viewed as a simple probabilistic summation over an orthonormal set of well prepared eigenstates in a somewhat classical scenario.
 \end{conclusion}
\section{The Microscopic Structure of the Black hole Singularity}
Beginning with the fundamental paper \cite{Bardeen} which formally relates the behavior of the BH event horizon to the well known laws of thermodynamics, it seemed to be more or less evident that it is the event horizon which is playing the essential role in this investigation. Only much later we find some sceptical remarks as to this widespread view. For the benefit of the reader we cite some brief remarks below, which can be found in the following papers:
\begin{quote}\cite{Unruh1}: Our belief is that information is lost in a black hole\ldots In quantum gravity, the classical singularity will undoubtedly be replaced by some other, weirder structure, but it should still act as a sink.    \\
\cite{Wald1}: In a footnote on p.22 (preprint) it reads: However, if any meaning can be given to the notion of where the entropy of a black hole resides, it seems much more plausible to me that it resides in the deep interior of the black hole (corresponding to the classical space-time singularity).   \\
\cite{Wald2}, At the end of sect. 6.1: On the other hand, if pure states evolve to mixed states in a fully quantum treatment of the gravitational field, then at least the aspect of the classical singularity as a place where information can get lost must continue to remain present in quantum gravity.
\end{quote}

It is evident that the event horizon seems to behave in a way similar to a thermodynamic system. This is, at first glance, not so clear in the case of the BH singularity. We will try to clarify this in the following investigation. We have seen that the entropy of the BH, associated with the event horizon is proportional tro the area $A$:
\beq S_{BH}=A/4L^2_p\quad \text{with}\quad  A=16\pi(GM/c^2)^2  \eeq
That is, it is proportional to $M^2$. We want to show that one can associate an entropy with the BH singularity which is proportional to $M^2$ and that the proportionality to $A$ is perhaps only a secondary effect.

To this end we have, in a first step, to clarify the presumed microscopic nature of the classical BH singularity. Put differently, we have to replace what is in the classical context a point singularity of the space-time manifold, $ST$, microscopically, that is, perhaps on Planck scale level, by an infinitesimal neighborhood. Usually it is in a somewhat cursory way stated that space and time are dissolved or destroyed at the BH singularity. We will try to give a more concrete meaning to this vague picture.
\begin{bem} Note that this vague picture necessarily implies that something more primordial is expected to exist into which this singularity is dissolved. In a first step we want to describe this presumed microscopic structure. We already introduced in \cite{Requ3} the notion of \tit{physical points}, i.e. points, having a rich internal structure under a microscope.
\end{bem}

We assume the following to hold:
\begin{conjecture} The local state of the infinitesimal, microscopic, Planck scale region, representing the classical BH singularity, is of the same nature as the state which prevailed befor the Big Bang and/or cosmic inflation happened.
\end{conjecture}
We briefly discussed the nature of this latter scenario in sect. 2 of \cite{Requ1} and we still have to be brief in the following as a more detailed discussion is quite complex and would go beyond the aims of the present paper. Some more remarks can be found in the papers we are citing below. We will only provide the concepts we will need in the followiing investigation.
\begin{bem} It is obvious that this topic, i.e., the phase transition, starting from some primordial initial state and unfolding ultimately into our present universe, should be closely connected with the inflationary universe scenario as it is e.g. beautifully described by A.Guth in \cite{Guth1} or in \cite{Kolb}.
\end{bem}
One should however note that in the following we will develop a picture in which space-time on its most fundamental level is no longer a smooth manifold while the socalled \tit{false vacuum} of the inflationary scenario from which our present universe is assumed to have emerged is, at least as far as we can see, an unfolding field theory on an expanding smooth space-time.

We now undertake to describe the nature of this primordial phase from which our present universe did emerge in a primordial phase transition, dubbed $PT$, which precedes all the following phase transitions which are described in the literature. The original primordial phase we denote by $QX$, the emergent \tit{(quantum) space-time} by $(Q)ST$, the emergent \tit{quantum matter} by $QM$. We assume that $QX$ is an irregular, wildly fluctuating substratum which does not support any stable structures and does not have a pronounced \tit{near/far order}. It is assumed to consist of a large network of elementary degrees of freedom which do interact in a very erratic way with each other. $ST$ denotes the smooth classical space-time and $M$ classical matter. To put it briefly:
\begin{conjecture} We assume that our present universe emerged from a first order phase transition
\beq QX\overset{PT}{\rightarrow}(Q)ST+(Q)M  \eeq
The latent heat is transformed (at least to a large part) into quantum matter $QM$. We assume, furthermore, that $QST$ represents a more ordered phase, that is, both its energy content $E(QST)$ and its entropy content $\hat{S}(QST)$ is lower compared to the phase $QX$. We have
\beq E(QX)=E(QST)+E(QM)\quad \text{and}\quad \hat{S}(QX)\leq \hat{S}(QST)+\hat{S}(QM)  \eeq
\end{conjecture}
\begin{bem}We described such a phase transition plus the necessary coarse graining and renormalization steps in more detail in e.g. \cite{Requ1} and further references given there.
\end{bem}

\subsection{The Graph Network of Elementary Interaction}
In a next step we want to make some remarks about the microscopic structure of the primordial phases $QX$ and $QST$. As the two approaches, \tit{string theory} and \tit{loop quantum gravity}, are widely discussed in the literature, we rather prefer to discuss a framework called \tit{pregeometry}. This point of view was  advocated by J.A.Wheeler, who discussed it in e.g. the monumental \cite{Wheeler}. This point of view coincides to a large extent with our own working philosophy, that is, to make only very few general and, as we think, natural assumptions concerning the structure and constituents of this primordial substratum.

An early and very interesting discussion one can find in \cite{MengerSchilpp}. In his essay Menger (a mathematician) introduced what one may call \tit{random metric spaces}. Points are no longer primary entities, instead we assume a geometry of \tit{lumps} which underlies our classical continuous space-time $ST$. An early contribution by Menger is \cite{Menger1}. A more detailed representation is \cite{Sklar}. Interesting are Einstein's replies in \cite{Schilpp}. He argues that he adheres to the continuum because he has been unable to think up anything organic to take its place. We related Menger's to our approach in \cite{RequRoy}.

The overall point of view concerning \tit{pregeometry} can be stated in a Wheelerian way: \tit{Geometry from Non-Geometry}. Before we discuss our own approach we want to mention a few sources which more or less strongly do belong to this working philosophy. A certain point of view is that cellular automata (CA) are a promising starting point. we mention for example the nice book by A.Ilachinski (\cite{Ila}) or `t Hooft's work (see e.g. \cite{HooftCA}). A short overview of some of the various approaches in the field of pregeometry is \cite{Meschini}. A related philosophy is developed in e.g. \cite{Trugenberger}, see also \cite{Finkelstein}. We furthermore mention \cite{Konopka} and our related paper \cite{RequRastgoo}. 

There are basically two different classes of dynamical laws in this field. While we mainly employed a generalized CA-law, some other colleagues use more Ising-like statistical laws. Furthermore we generalized the CA-approach to what we called: \tit{Structurally Dynamic Cellular Network} (SDCN), as we think, the pure CA-appoach is too rigid, given that the microscopic substratum of space-time should also behave dynamically.

In our view, the main task consists of developing a framework which starts from pregeometric entities like $QX$ and/or $QST$ which do not yet own any continuum concepts or notions and arrives, after the application of some sort of \tit{geometric renormalisation process} at a continuous space-time $ST$ carrying a definite dimension. This includes the emergence of notions like a \tit{differential structure}, \tit{dimensional concepts}, \tit{distance concepts}, \tit{Dirac operator} etc. In this context we cite a few of our papers in which we developed such a program, \cite{Requ2},\cite{Requ3},\cite{Requ4},\cite{Requ5},\cite{Requ6}.

 We assume that our networks $QX$ and $QST$ consist of elementary, irreducible degrees of freedom (DoF) which interact by means of local elementary interactions or \tit{information channels}, depending on the kind of models we have in mind. The local state spaces can, for example, be (finite dimensional) Hilbert spaces or (logical) gates. We also assume that the local interactions or channels can occupy different states or can even be switched on or off so that the \tit{wiring diagram} of local interactions is no longer a rigid underlying structure but may behave in a dynamical way.

For convenience (while this is not really a crucial assumption) we assume that our elementary DoF, denoted by $x_i$, are a countable set. The local interactions or channels we denote by $e_{ij}$ or $d_{ij}$, connecting the DoF $x_i$ and $x_j$, with $d_{ij}$ carrying, if necessary, a kind of direction (see below). All this now suggests the following mathematical model structure: We associate the elementary DoF with the nodes or vertices of a Graph, $G$, the elementary interactions or channels with the edges of the graph $G$.
\begin{defi} A simple, countable, labeled, undirected graph, G, consists of a countable set of nodes or vertices, V, and a set of edges or bonds, E, each connecting two of the nodes. There exist no multiple edges (i.e., edges connecting the same pair of nodes) or elementary loops (a bond starting and ending at the same node). In this situation the edges can be described by giving the corresponding set of unordered pairs of nodes. The members of V are denoted by $x_i$, the edges by $e_{ij}$, connecting the nodes$x_i$ and $x_j$.
\end{defi}

A large class of possible models are our structurally dynamic cellular networks (SDCN). the assumed dynamical evolution laws have the following general structure:
\begin{defi} (General local law on networks). Each node can be in a number of internal states $s_i \in \mcal{S}$. Each bond $e_{ik}$ carries a corresponding bond state $J_{ik}\in \mcal{J}$. Then the following general transition law is assumed to hold:
\begin{align} s_i(t+\tau)&=ll_x(\{s_k(t)\},\{J_{kl}\})\\
J_{ik}(t+\tau)&=ll_J(\{s_l(t)\},\{J_{lm}(t)\})\\
(\underline{S},\underline{J})(t+\tau)&=LL(\underline{S},\underline{J})(t)
\end{align}
where $ll_x$, $ll_J$ are two maps (being the same over the whole graph) from the state space of a local neighborhood of the node or edge on the l.h.s. to $,\mcal{S},\mcal{J}$, yielding the updated values of $s_i$ and $J_{ik}$. $\mathcal{S}$ and $\mathcal{J}$ denote the global states of the nodes and edges  and LL the global law built from the local laws at each node and bond. \end{defi}
We performed large scale computer simulations to study the behavior of such networks and to search for phase transitions under varying parameter conditions in \cite{Nowotny}. It is wellknown in the field of random graphs that socalled \tit{threshold functions} (i.e. phase transitions) do exist (see below and e.g. \cite{Bollo2}).
\begin{bem}Simple cellular network laws with a switch on/off mechanism of bonds are, for example, formulated in e.g. \cite{Requ3} or \cite{Requ5}.
\end{bem}

As we are dealing in this field with very large networks or graphs, which are changing their shape during their dynamical developement, we expect the dynamics to be sufficiently stochastic so that a point of view may be appropriate which reminds us of the working philosophy of \tit{statistical mechanics}. It has in particular turned out that this approach is extremely efficient if we want to study socalled \tit{graph properties}. We discussed these topics in considerably more detail in e.g. \cite{Requ3} or \cite{Requ5}. In the following we will only briefly introduce the concept of \tit{random graph} for the convenience of the reader. 
\begin{defi}(The random graph idea) Take all possible labelled graphs over n nodes as probability space $\mcal{G}$ (that is, each graph represents an elementary event in this space). Give each edge the independent probability $0\geq p\leq 1$ (i.e., p is the probability that there exists a bond between the two vertices under discussion. Let $G_m$ be a graph over the above vertex set V having m bonds. Its probability is then
\beq pr(G_m)=p^m\cdot q^{N-m} \quad \text{with}\quad N=\binom{n}{2}\eeq
the maximal possible number of edges in the graph and $q=1-p$. There exist $\binom{N}{m}$ different labelled graphs $G_m$ having m bonds and the above probability is correctly normalized:
\beq pr(\mcal{G}=\sum_{k=0}^N\, \binom{N}{m}p^mq^{N-m}=(p+q)^N=1  \eeq
We have for example 
\beq pr(m)=\binom{N}{m}p^mq^{N-m}\quad \text{and}\quad <m>=\sum\, m\cdot pr(m)=Np \eeq
with $pr(m)$ the number of graphs with exactly m edges and $<m>$ the corresponding probabiliy.
This probability space is frequently denoted by $\mcal{G}(n,p)$.
\end{defi}

While this space is a good candidate for numerical computations, another interesting example is a \tit{microcanonical} version: 
\begin{bem} We consider graphs over n nodes but with a fixed number of m edges, each graph having the probability $\binom{N}{m}^{-1}$ as there are exactly $\binom{N}{m}$ of them. The corresponding probability space is denoted by $\mcal{G}(n,m)$.
\end{bem}
A comprehensive account of graph theory is e.g. \cite{Bollo1}, random graphs are treated in quite some detail in \cite{Bollo2}

We will now provide a few concepts, being useful in the treatment of graph properties.
\begin{lemma}We assume for convenience that all our graphs or networks, we are employing, are connected, i.e., each pair of vertices, $x_I,x_j$, can be connected by an edge sequence, starting at $x_i$ and ending at $x_j$. The length of the edge sequence is given by the number of edges in the sequence. The minimal number of edges in the edge sequences connecting $x_i$ and $x_j$ defines a distance function on the graph and, by the same token, a metric. Thus graphs are discrete metric spaces (cf. e.g. \cite{Requ3} or \cite{Requ5}).
\end{lemma}

To perform certain renormalisation steps, starting from some discrete, erratic network structure, and ending up, hopefully, at some smooth manifold, we employ a certain graph theoretic concept, introduced in the following definition.
\begin{defi}(Subsimplices and cliques) With G a given fixed graph and $V_i$ a subset of its vertex set V, the corresponding induced subgraph over $V_i$ is called a subsimplex $S_i$ or complete subgraph if all its vertex pairs are connected by bonds living in E (the edge set of G). In this class there exist certain maximal subsimplices, that is, every further addition of a node distroys this property. These maximal subsimplices are called cliques and are the canditates for our physical points, i.e. classical points of our assumed manifold, having an internal structure.
\end{defi}
\begin{bem} Note that the transition to a smooth manifold is a quite complex process with quite an amount of modern mathematics being needed. This is described in more detail in our above mentioned papers (see e.g. \cite{Requ6}).
\end{bem}
\subsection{The Entanglement network}
We will now introduce an, as we think, important generalization. E.Schroedinger in 1935 made a crucial observation (\cite{Schroedinger}):
\begin{quote} I would not call that one but the characteristic trait of quantum mechanics, the one that enforces its entire departure from classical lines of thought.
\end{quote}
That is, instead of employing elementary interactions or information channels, we want to view the edges of the graph as certain entanglement relations which we will explain below.
\begin{conjecture}While the phenomenon of entanglement, which is closely related to the property of complex linearity,  is not really well understood ontologically on the present level of quantum physics, we think, it represents a kind of primordial interaction like relation on the level of pregeometry and corresponding more microscopic structures underlying our present day quantum theory.
\end{conjecture}
\begin{bem}The notion of entanglement can be succinctly expressed as follows: Two quantum systems one Hilbert spaces $\mcal{H}_1,\mcal{H}_2$, with joint Hilbert space the tensor product $\mcal{H}_1\otimes \mcal{H}_2$, and joint state $\omega_{12}$ are not entangled and called separable if $S_1$ and $S_2$ have their own independent states $\omega_1$ and $\omega_2$ with
\beq \omega_{12}=\omega_1\otimes \omega_2  \eeq
with the $\omega_i$ some density matrices. They are entangled if this does not hold.
\end{bem}
This property played a crucial role in the early debates around the EPR-paper \cite{Einstein}.

In the present discussion of entanglement (e.g. in quantum information theory) a slightly more general definition of separability is used:
\begin{defi}A bipartite state $\omega_{12}$ is separable if it can be written as
\beq \omega_{12}=\sum_i\, p_i\cdot \omega_1^i\otimes \omega_2^i\quad ,\quad p_i\geq 0 \; ,\; \sum\, p_i=1  \eeq
This implies that there do exist only classical correlations.
\end{defi}
As the literature about entanglement is extremely large, we will mention only two representative papers (\cite{Horodecki} and \cite{Casini}). The typical fields of application of entanglement properties are quantum information theory, quantum many body physics and, to some extent, BH physics. But recently it was also applied in quantum gravity or quantum space-time physics, more specifically, in the subfield of \tit{gauge theory/gravity duality} (see e.g. \cite{Swingle} and \cite{Raamsdonk}).

While, in our view, it is not guaranteed that quantum theory necessarily is holding sway on the most primordial level, e.g. on the Planck scale, we assume, employing our picture of a \tit{geometric renormalization group}, that this may hold at least on some intermediate level. That is, we assume that we have some \tit{tensor network} living on a corresponding \tit{simple graph} as described above. But as the notion of entanglement, in particular in such \tit{multipartite structures} as a tensor network, is a relatively delicate and subtle property (cf. e.g. the section on multipartite entanglement in \cite{Horodecki}), we have, in a first step, to develop and explain various points.

We assume that we are given a large tensor product of $N$ Hilbert spaces, $\mcal{H}_i, i=1,\ldots ,N$.
 For convenience we furthermore assume that all the Hilbert spaces are finite dimensional. That is:
 \begin{defi}We have
 \beq \mcal{H}:=\bigotimes_i\,\mcal{H}_i=\mcal{H}_1\otimes\cdots\otimes\mcal{H}_N  \eeq
with monomials $\bigotimes_i\psi_i\; ,\;\psi_i\in \mcal{H}_i$. If $\{e^i_{\nu}\}$ is a basis in $\mcal{H}_i$ a general Hilbert space vector $\Psi\in \mcal{H}$ can be represented as 
\beq \Psi=\sum\, c^{\nu_1\ldots\nu_N}\cdot e^1_{\nu_1}\otimes\cdots e^N_{\nu_N}  \eeq
\end{defi}

On $\mcal{H}=\bigotimes_i\mcal{H}_i$ we are given the tensor algebra $\mcal{A}$ with monomials 
\beq A_1\otimes\cdots\otimes A_N\; ,\; A_i\in \mcal{A}_i \;\text{the algebra of operators on}\; \mcal{H}_i     \eeq
We assume now that we have (for simplicity reasons) a pure state $\Psi$ on $\mcal{H}$ and a subcomplex of Hilbert spaces
\beq \mcal{H}_S=\bigotimes\mcal{H}_j\; ,\; S\subset \{1,\ldots ,N\}\; ,\; j\in S   \eeq
with corresponding tensor subalgebra $\mcal{A}_S$.
\begin{ob}The reduced state of $\Psi$ relative to $(\mcal{H}_S,\mcal{A}_S)$ is defined by 
\beq \omega_S(A_S):=(\Psi|A_S\otimes\mbf{1}_{S'}\Psi)=Tr\, (\rho_S\cdot A_S)  \eeq
$S'$ the complement of $S$ in $\{1,\ldots ,N\}$ and with $\rho_S$ the reduced density matrix on $(\mcal{H}_S,\mcal{A}_S)$ given by (for ease of notation we use indices $(12),(1),(2)$)
\beq (\psi|\rho_1\phi)=\sum_i\,(\psi\otimes e_i|\rho_{12}\phi\otimes e_i)  \eeq
with $\rho_{12},\rho_1,\rho_2$ density matrices on $\mcal{H}_{12},\mcal{H}_1,\mcal{H}_2$, $\psi,\phi$ vectors in $\mcal{H}_1$ and $\{e_i\}$ an orthonormal basis of unit vectors in $\mcal{H}_2$ ($\Psi$ being expressed as $|\Psi><\Psi|=\rho_{12}$ in our case). I.e., a pure state on $\mcal{H}$ becomes (in general) a mixed state on $\mcal{H}_S$.
\end{ob}

We can go one step further by taking a subset $U$ of $S$ and construct a corresponding mixed state on $U$:
\begin{ob} \beq \omega_U(A_U)= \omega(A_U\otimes \mbf{1}_{U'})=\omega_S(A_U\otimes\mbf{1}_{U''}) =Tr(\rho_U\cdot A_U)  \eeq
with $U',U''$ the complements of $U$ in $\{1,\ldots ,N\},S$, $\omega_U$ the reduced state relative to $\omega_S$, $\rho_U$ the reduced density matrix relative to $\rho_S$ or $\omega$.
\end{ob}

We said above that entanglement properties happen to be more complicated in multipartite systems. We would like to discuss at least one, in our view, particularly important property in this context.
\begin{defi} Let $\mcal{H}=\bigotimes_i\,\mcal{H}_i$ be some tensor product or some sub tensor product $\mcal{H}_S$ contained in a larger product, $\omega$ a state on $\mcal{H}$ or $\mcal{H}_S$. $\omega$ is called maximally entangled if the following holds:\\
For any bipartite decomposition, $S=U\cup U'$, $U'$ the complement of $U$ in $S$, 
\beq \omega\neq \sum\, p_i\cdot \omega_U^i\otimes\omega_{U'}^i   \eeq
\end{defi}

\begin{bem} We note that this property is closely related to our notion of \tit{complete subgraph} and \tit{cliques} we defined above in graph theory. 
\end{bem}

We will now discuss the relation of our graph theoretic approach, introduced above, to the concept of \tit{separability} and \tit{entanglement} of multipartite systems. 
\begin{bem}While it is not really crucial, we assume in the following that the notion of separability is adequately described on our primordial level we are envisaging by
\beq \omega=\omega_U\otimes\omega_{U'}  \eeq
as we think, the sum over probabilities, $p_i$, is rather signalling a classical behavior which should be absent on the presumed fundamental level.
\end{bem}

We introduced above the notion of a \tit{simple graph} with vertices $V=\{x_i\}$ and simple edges $E=\{e_{ij}\}$, connecting the vertices $x_i$ and $x_j$. We now want to relate this picture to corresponding entanglement properties of a \tit{tensor network}. We associate a Hilbert space $\mcal{H}_i$ to each vertex $x_i$. Important is the distribution of edges as it encodes the pregeometric structure of our system.

We assume a state, $\omega$, be given on $\bigotimes_i\,\mcal{H}_i$. We select two arbitrary vertices $x_i,x_j$ in the vertex set $V$ and define the relative state $\omega_{ij}$ on $\mcal{H}_i\otimes\mcal{H}_j$ induced by the fixed given global state $\omega$.
\beq \omega_{ij}(A_{ij})=\omega (A_{ij}\otimes\mbf{1}_{S'})  \eeq
$S'$ the complement of $\{i,j\}$ in $V$.
\begin{defi}i) If $\omega_{ij}$ is separable, i.e., if there exist two states, $\omega_1,\omega_2$, so that $\omega=\omega_1\otimes\omega_2$, the edge between $x_i,x_j$ is empty.\\
ii) If $\omega_{ij}$ is entangled, we connect $x_i,x_j$ by an edge.
\end{defi}
\begin{conclusion}In this sense we relate an entanglement graph, $G_e$, to our state $\omega$ on the tensor product $\bigotimes_i\mcal{H}_i$. When we assume a kind of dynamics to exist on our system,  which constantly is changing the state $\omega$, this will change, by the same token, the geometry of our entanglement graph.
\end{conclusion}

While we restrict our investigation (for the time being) to simple graphs, one should note that we can introduce more complex graph structures if it would become necessary. We will now show that our above construction of an entanglement graph with the help of pairs of vertices is consistent in a sense we will now prove. We start with a simple lemma:
\begin{lemma}The existence of entanglement in a tensor product $\mcal{H}_1\otimes\mcal{H}_2$ can already be proved on monomials $A_1\otimes A_2$.
\end{lemma}
Proof: Due to linearity, if a state is separable on all products like $A_1\otimes A_2$ it is separable on the full tensor product $\mcal{H}_1\otimes\mcal{H}_2$. This implies that in case $\omega$ is not separable, i.e., is entangled, this can already be exhibited on products like $A_1\otimes A_2$.

We now prove the following:
\begin{satz}We divide the vertex set $V$ of the graph $G_e$, defined by a global state $\omega$, into $S\cup S'$ and assume that there exists a pair $x_i\in S,x_j\in S'$ so that the reduced state $\omega_{ij}$ is entangled on $\mcal{H}_i\otimes\mcal{H}_j$, that is
\beq \omega_{ij}\neq \omega_i\otimes\omega_j  \eeq
That is, there exists an edge connecting $x_i,x_j$. Then $\omega$ is entangled relative to $\mcal{H}_S,
\mcal{H}_{S'}$, i.e.
\beq \omega\neq\omega_S\otimes\omega_{S'}  \eeq
\end{satz}
Proof: We have that
\begin{align} \omega(A_i\otimes \mbf{1})&=\omega_{ij}(A_i\otimes\mbf{1})=\omega_i(A_i)=\omega_S(A_i\otimes\mbf{1})\\
\omega(\mbf{1}\otimes B_j)&=\omega_{ij}(\mbf{1}\otimes B_j)=\omega_j(B_j)=\omega_{S'}(B_j\otimes\mbf{1})
\end{align}
with $\mbf{1}$ acting on the respective dual spaces. By assumption there exist pairs, $A_i,B_j$, so that
\beq \omega_{ij}\neq
\omega_i(A_i)\cdot\omega_j(B_j)  \eeq
This implies
\beq \omega(A_i\otimes B_j\otimes\mbf{1})\neq\omega_S(A_i)\cdot\otimes\mbf{1})\cdot\omega_{S'}(B_j\otimes\mbf{1})  \eeq
with $\mbf{1}$ the unit operators on the respective dual spaces. I.e., if $\omega_{ij}$ is entangled, $\omega$ is also entangled relative to $S,S'$.
\begin{koro} If the entanglement graph $G_e$ is maximally entangled according to our definition via pairs of Hilbeert spaces, $\mcal{H}_i,\mcal{H}_j$, then $\omega$ on $\mcal{H}$ is maximally entangled according to the definition of entanglement of states, that is, $\omega$ is entangled relative to any bipartite division of $\mcal{H}$ into $\mcal{H}_S,\mcal{H}_{S'}$.
\end{koro}
Proof: As above, we select in any bipartite division $S,S'$ a pair $x_i\in S,x_j\in S'$ and procede in the same way as above in the proof of the theorem.

We want to close this section by showing that one can, in a natural way, attribute a value to the edges of $G_e$ in form of a strength of the respective connection. We will employ the notion of entropy to define such a strength of the connection. The \tit{v.Neumann entropy} of $\rho$ is given by
\beq S(\rho):= -Tr(\rho\cdot \ln\,\rho) \eeq
\begin{satz}We have the important sub-additivity inequality for the relative entropy:
\beq S(\rho_{12})\leq S(\rho_1)+S(\rho_2)=S(\rho_1\otimes \rho_2)  \eeq
\end{satz}
(More on this important topic can e.g. be found in \cite{Lanford},\cite{Araki} or \cite{Ruelle}).

It is then natural to compare the lhs and the rhs of the above equation and define:
\begin{defi}(Mutual Information)
\beq I(\mcal{H}_1\otimes\mcal{H}_2):=S(\rho_1)+S(\rho_2)-S(\rho_{12})\geq 0  \eeq
\end{defi}
\begin{koro} Note that $I(\mcal{H}_1\otimes\mcal{H}_2)$ exactly vanishes if $\rho_{12}=\rho_1\otimes\rho_2$, i.e., no edge according to our definition.
\end{koro}
\begin{conclusion}We now attribute such a notion to the reduced states $(\omega_{ij},\rho_{ij})$ and $(\omega_i,\rho_i),(\omega_j,\rho_j)$.
\end{conclusion}
\section{The Entropy of the Singularity}
The preceding sections represented the necessary steps to be able to motivate in this final section our claim that the BH entropy is located in the BH singularity or, to state it in a slightly different form, that the localization of the entropy is observer dependent, i.e., is after all a consequence of the equivalence of coordinate systems, that is, either a static reference system outside of the BH horizon or, for example, a freely falling observer.

We argued that the fine structure of $QX$ and $QST$ consists of a large array of elementary DoF which are related or connected by elementary interactions, information channels or entanglement relations. We came to the conclusion that we can represent the \tit{wiring diagram} by a huge graph of, say $n$ vertices being connected by a number of $N'$ edges.
\begin{bem}In the following we will usually only deal with some subgraph of the total graph, for example with the subset of $QX$ that represents the infinitesimal region belonging to the macroscopic BH singularity. That is, in that case the parameters $n$ and $N'$ belong to the corresponding subgraph.
\end{bem}
\begin{ob}The maximal possible number of edges is
\beq n\cdot(n-1)/2=\binom{n}{2}=:N  \eeq
\end{ob}
\begin{assumption}We assume that $QX$ can be approximated by a kind of equilibrium state and that the primordial dynamics (among other things) is constantly switching the various edges on and off. We hence expect that the state of $QX$ is a state of maximal entropy.
\end{assumption}

We will concentrate ourselves in the following on the \tit{configuration entropy} of our graph state as we expect this to be the main contribution. We now show the following:
\begin{ob}The numberof different graph configurations, consisting of exactly $N'$ edges is $\binom{N}{N'}$, that is the number of $N'$-sets in a set of $N$ elements.
\end{ob}
\begin{lemma}It holds
\beq \binom{n}{r}=(n-r+1)/r\cdot\binom{n}{r-1} >(n/r-1)\cdot\binom{n}{r-1}  \eeq
\end{lemma}
Proof: This follows directly from the definition 
\beq \binom{n}{r}=n!/r!\cdot (n-r)!  \eeq
We therefore have for $r<n/2$:
\beq n/r-1>1\;\text{that is}\;\binom{n}{r}>\binom{n}{r-1}  \eeq
\begin{conclusion}The maximum of $\binom{n}{r}$ is attained for $r=n/2$ (where, for simlicity reason, we assume $n$ to be even.
\end{conclusion}

In a next step we estimate the quantity $\binom{N}{N/2} =N!/(N/2)!(N/2)!$ or, rather, its logarithm for large $N$. We use Stirling's formula
\beq N!\approx \sqrt{2\pi N}\cdot (N/e)^N  \eeq
We have ($\ln e=1$):
\begin{align}
\ln N! & \approx \ln\sqrt{2\pi N}+N\ln N-N\\
\ln (N/2)!^2 & \approx 2[\ln\sqrt{2\pi N/2}+N/2\ln N/2-N/2]  
\end{align}
Hence
\begin{multline} \ln\binom{N}{N/2}\approx [\ln\sqrt{2\pi N}+N\ln N-N]-\\
2[\ln\sqrt{2\pi N/2}+N/2\ln N/2-N/2]=\ln\sqrt{2\pi N}-2\ln\sqrt{\pi N}+N\ln 2\\
\approx  N\ln 2\quad (+1/2\cdot\ln N)
\end{multline}
\begin{conclusion}The maximum of configuration entropy is attained for $N'$, the numberof edges in the graph, equal to $N/2$ and has the value
\beq \ln\binom{N}{N/2}\approx N\ln 2\;\text{with}\; N=\binom{n}{2}=n\cdot (n-1)/2\approx n^2/2  \eeq
\end{conclusion}

Up to now we assumed that the graph has exactly $\binom{N}{N/2}$ edges. In reality this number will fluctuate around some mean value. The following result will show that the final result will be essentially unaffected by such a restriction. 
\begin{lemma}It holds
\beq \sum_k \binom{N}{k}=2^N\quad \text{i.e.}\quad \ln \biggl(\sum_k \binom{N}{k}\biggr)=N\cdot\ln 2  \eeq
that is, we get the same result if the number of edges is allowed to vary.
\end{lemma}
Proof: A result following from the binomial formula for
\beq (x+y)^N=\sum_k \binom{N}{k}x^ky^{N-k}  \eeq
with $x=y=1$.

We now proceed to establish the functional dependence of the above entropy on the mass of the BH. We think it is natural to relate the macroscopic mass, $M$, of the BH to the number of microscopic DoF and write:
\begin{conclusion}We assume that 
\beq M=n\cdot M_p\quad ,\quad M_p=\sqrt{\hbar c/G}  \eeq
holds where $n$ now belongs to the subgraph, representing the infinitesimal region which is associated with the macroscopic BH singularity. This implies by the same token that $M_p\cdot c^2$ is the energy necessary to transform a Planck cell of size $l_p^3$ located in $QST$ into the corresponding cell in $QX$.
For the entropy it follows that 
\beq \hat{S}\approx N\ln 2\approx (\ln 2/2)\cdot n^2=(\ln 2/2)\cdot M^2/M_p^2  \eeq
\end{conclusion}

We now want to estimate the extension of the infinitesimal region which macroscopically is considered as the BH singularity. Again we invoke the fundamental role played by the system of Planck units.
\begin{assumption}We associate an elementary DoF with the corresponding Planck volume
\beq l_p^3=(\sqrt{\hbar G/c^3})^3\approx 10^{-105}m^3  \eeq
\end{assumption}
To show that the region being associated to the BH singularity is indeed infinitesimal on the macroscopic level we take for example the mass of our sun
\beq M_{sun}\approx 10^{53}g\quad\text{with}\quad M_{sun}/M_p\approx 10^{38}  \eeq
We have
\beq 10^{38}\cdot l_p^3\approx 10^{-67}m^3  \eeq
for the volume of the infinitesimal region and hence for the presumed radius of the infinitesimal region associated to the BH singularity:
\beq R_{sing}(sun)\approx 10^{-22}m  \eeq
\begin{conclusion}The extension of the infinitesimal region associated to the BH singularity is indeed infinitesimal in SI-units.
\end{conclusion}

\end{document}